\documentclass[twocolumn,showpacs,preprintnumbers,amsmath,amssymb,pre]{revtex4}

\usepackage{times,graphicx,amssymb,amsmath}

\newcommand{\avg}[1]{\langle{#1}\rangle}
\begin{document}

\title{ Phenomenological Models of Socio-Economic Network Dynamics }

\author{George C.M.A. Ehrhardt}
\email{gehrhard@ictp.trieste.it}
\author{Matteo Marsili}
\affiliation{The Abdus Salam ICTP, Strada Costiera 11, I-34014, Trieste. Italy.}
\author{Fernando Vega-Redondo}
\affiliation{Universidad de Alicante, Facultad de Economicas, Universidad de Alicante, 03071, Alicante. Spain.}
\altaffiliation[also ]{University of Essex, Wivenhoe Park, Colchester, CO4 3SQ, UK.}

\begin{abstract}
We study a general set of models of social network evolution and dynamics.  
The models consist of both a dynamics on the network and evolution of the network.
Links are formed preferentially between 'similar' nodes, where the similarity is defined by the particular process taking place on the network.
The interplay between the two processes produces phase transitions and hysteresis, as seen using numerical simulations for three specific processes.
We obtain analytic results using mean field approximations, and for a particular case we derive an exact solution for the network.
In common with real-world social networks, we find coexistence of high and low connectivity phases and history dependence.  
\end{abstract}


\maketitle


\section{Introduction}
\label{sec:intro}

In recent years physicists have paid much attention to network
structures -- describing either technological infrastructures or
biological, genetic, logical or social relationships -- as they
play a prominent role in shaping the nature of the processes
taking place on them and the resulting collective behavior.
Examples of how the structure affects the function of networked
systems include the importance of shortcuts in endowing finite
dimensional networks of the small world property
\cite{WattsStrogatz} and of scale-free degree distribution for
robustness against failure \cite{Cohen} or the relevance of motifs
for specific dynamical properties \cite{motifs}.

Socio-economic networks offer an example where the relation
between structure and function is not unidirectional. Indeed their
structure is inherently dynamical and it is shaped by the
incentives of agents, i.e. by the socio-economic functions
provided by the network. This paper discusses a class of generic
model of stochastic dynamical social networks which make the
interplay between structure and function of social network
explicit in a simple way. We consider a set of agents -- be they
individuals or organisations -- who establish bilateral
interactions (links) when profitable. The network evolves under
changing conditions. That is, the favourable circumstances that
led at some point to the formation of a particular link may later
on deteriorate, causing that link's removal.  Hence volatility
(exogenous or endogenous) is a key disruptive element in the
dynamics.  Concurrently, new opportunities arise that favour the
formation of new links. Whether linking occurs depends on factors
related to the similarity or proximity of the two parties. For
example, in cases where trust is essential in the establishment of
new relationships (e.g. in crime or trade networks), linking may
be facilitated by common acquaintances or by the existence of a
chain of acquaintances joining the two parties. In other cases
(e.g. in R\&D or scientific networks), a common language,
methodology, or comparable level of technical competence may be
required for the link to be feasible or fruitful to both parties.

In a nutshell, our model conceives the dynamics of the network as
a struggle between volatility (that causes link decay) on the one
hand, and the creation of new links (that is dependent on
similarity) on the other. The model must also specify the dynamics
governing inter-node similarity.  A reasonable assumption in this
respect is that such similarity is enhanced by close interaction,
as reflected by the social network. For example, a firm (or
researcher) benefits from collaborating with a similarly advanced
partner, or individuals who interact regularly tend to converge on
their social norms and other standards of behavior.

We study different specifications of the general framework, each
one embodying alternative forms of the intuitive idea that
\textquotedblleft interaction promotes
similarity.\textquotedblright\ Our main finding is that in all of
these different cases the network dynamics exhibits a rich
phenomenology characterized by a) sharp phase transition b)
resilience, i.e. stability against deteriorating conditions and c)
equilibrium coexistence. The essential mechanism at work is a
positive feedback between link creation and inter-node similarity,
these two factors each exerting a positive effect on the other.
Feedback forces of this kind appear to operate in the dynamics of
many social networks. We show that they are sufficient to produce
the sharp transitions, resilience, and equilibrium co-existence
that, as we will discuss in the next section, are salient features
of many socio-economic phenomena. Finally, this phenomenology
bears a formal similarity with the liquid-gas phase transition,
thus suggesting that a classification in terms of phases may be
applicable also to socio-economic networks.

The rest of this paper is organized as follows: The next section
discusses in an introductory way the empirical evidence which our
model addresses. In section \ref{sec:model} we outline the general
setup and a generic model of which we will discuss particular
realizations the following sections. In particular, we shall first
discuss the case where network formation depends on the topology
of the network (sect. \ref{sec:chain}) and then cases where it is
coupled with the dynamics of a continuous (sect. \ref{sect:h}) or
discrete (sect. \ref{sec:potts}) variable. These two models
addresses situations where homogeneity in some dimension (e.g.
technological levels or knowledge) or coordination (on e.g.
a standard) play a crucial role, respectively. Numerical simulations
will be supplemented by mean field analysis which provides a
correct qualitative picture in all cases and, in some cases, accurate
quantitative estimates. A case where an exact solution
can be derived will be described in section \ref{sec:exact}.
In section \ref{sec:concl} we end with some concluding remarks.

\section{Empirical stylized facts of socio-economic networks}
\label{sec:facts}

There is a growing consensus among social scientists that many
social phenomena display an inherent network dimension. Not only
are they \textquotedblleft embedded\textquotedblright\ in the
underlying social network \cite{Granov} but, reciprocally, the
social network itself is largely shaped by the evolution of those
phenomena. The range of social problems subject to these
considerations is wide and important. It includes, for example,
the spread of crime \cite{Glaeseretal, Haynie} and other social
problems (e.g. teenage pregnancy \cite{Crane, Harding}), the rise
of industrial districts \cite{OECD, Saxenian, Granovetteretal},
and the establishment of research collaborations, both scientific
\cite{Newman, Goyaletal} and industrial \cite{Hagedoorn,Kogut}.
Throughout these cases, there are a number of interesting
observations worth highlighting:\medskip

(a) \textbf{Sharp transitions}:\emph{\ The shift from a sparse to
a highly connected network often unfolds rather \textquotedblleft
abruptly,\textquotedblright\ i.e. in a short timespan}. For
example, concerning the escalation of social pathologies in some
neighborhoods of large cities, Crane \cite{Crane} writes that
\textquotedblleft ...if the incidence [of the problem] reaches a
critical point, the process of spread will
explode.\textquotedblright \ Also, considering the growth of
research collaboration networks, Goyal \emph{et al.} \cite{Goyaletal} report a steep increase in the per capita number of
collaborations among academic economists in the last three
decades, while Hagerdoorn \cite{Hagedoorn} reports an even sharper
(ten-fold) increase for R\&D partnerships among firms during the
decade 1975-1985.\medskip

(b)\textbf{\ Resilience}:\emph{\ Once the transition to a highly
connected network has taken place, the network is robust,
surviving even a reversion to \textquotedblleft
unfavorable\textquotedblright\ conditions}. The case of
California's Silicon Valley, discussed in a classic account by
Saxenian \cite{Saxenian}, illustrates this point well. Its
thriving performance, even in the face of the general crisis
undergone by the computer industry in the 80's, has been largely
attributed to the dense and flexible networks of collaboration
across individual actors that characterized it. Another
intrinsically network-based example is the rapid recent
development of Open-Source software (e.g. Linux), a phenomenon
sustained against large odds by a dense web of collaboration and
trust \cite{Benkler}. Finally, as an example where
\textquotedblleft robustness\textquotedblright\ has negative
rather than positive implications,\ Crane \cite{Crane} describes
the difficulty, even with vigorous social measures, of improving a
local neighborhood once crime and other social pathologies have
taken hold. \medskip

(c) \textbf{Equilibrium co-existence}: \emph{Under apparently
similar environmental conditions, social networks may be found
both in a dense or sparse state}. Again, a good illustration is
provided by the dual experience of poor neighborhoods in large
cities \cite{Crane}, where neither poverty nor other
socio-economic conditions (e.g. ethnic composition) can alone
explain  whether or not there is degradation into a ghetto with
rampant social problems. Returning to R\&D partnerships, empirical
evidence \cite{Hagedoorn} shows a very polarized situation, almost
all R\&D partnerships taking place in a few (high-technology)
industries. Even within those industries, partnerships are almost
exclusively between a small subset of firms in (highly advanced)
countries \cite{footnote1}.

From a theoretical viewpoint, the above discussion raises the
question of whether there is some common mechanism at work in the
dynamics of social networks that, in a wide variety of different
scenarios, produces the three features explained above: (a)
discontinuous phase transitions, (b) resilience, and (c)
equilibrium coexistence. Our aim in this paper is to shed light on
this question within a general framework that is flexible enough
to accommodate, under alternative concrete specifications, a rich
range of social-network dynamics.

\section{The model}
\label{sec:model}

Consider a set ${\cal N }=\{1,\ldots ,n \}$ of agents, whose state
and interactions evolve in continuous time $t$.  They form the
nodes of a network which is described by a non-directed graph
$g(t)\subset \{ij:i\in {\cal N },~j\in {\cal N }\}$, where
$ij(\equiv ji)\in g(t)$ iff a link exists between agents $i$ and
$j$. The network evolution is modelled in terms of continuous time
stochastic elementary Poisson processes, and it is therefore
defined by specifying the rates at which these processes
occur \cite{footnote2}.  Firstly, each node $i$ receives an
opportunity to form a link with a node $j$, randomly drawn from
${\cal N }$ ($i \ne j$), at rate $\eta$. If the link $ij$ is not
already in place, it forms with probability

\begin{equation}
P\{ij\rightarrow g(t)\}=\left\{
\begin{array}{cc}
1         & \;\textrm{if } d_{ij}(t)\leq \bar{d} \\
\epsilon  & \;\textrm{if } d_{ij}(t) >   \bar{d}
\end{array}
\right.   \label{Pform}
\end{equation}
where $d_{ij}(t)$ is the \textquotedblleft
distance\textquotedblright\ (to be specified later) between $i$
and $j$ prevailing at $t.$  Thus if $i$ and $j$ are close, in the
sense that their distance is no higher than some given threshold
$\bar{d}$, the link forms at rate $\eta$; otherwise, it forms at a
much smaller rate $\eta \epsilon$. Secondly, each existing link
$ij\in g(t)$ decays at rate $\lambda$.  That is, each link in the
network disappears with probability $\lambda dt$ in a time
interval $[t,t+dt)$. We shall discuss three different
specifications of the distance $d_{ij}$, each capturing different
aspects that may be relevant for socio-economic interactions.

In all three cases, we mostly focus on the stationary state
behavior, which we shall illustrate using both numerical
simulations and a mean-field analytic approach. Concerning the
latter, we focus on the limit $n\rightarrow \infty $, for which
the analysis is simpler. We characterise the long run behavior of
the network solely in terms of the stationary degree distribution
$p(k)$, which is the fraction of agents with $k$ neighbours.  This
corresponds to neglecting degree correlations, i.e. to
approximating the network with a random graph (see
\cite{randomGraph}), an approximation which is reasonably accurate
in the cases we discuss here. The degree distribution satisfies a
master equation, which is specified in terms of the transition
rates $w(k\rightarrow k\pm 1)$ for the addition or removal of a
link, for an agent linked with $k$ neighbours. While
$w(k\rightarrow k-1)=\lambda k$ always takes the same form, the
transition rate for the addition of a new link
\[
w(k\rightarrow k+1)=2\eta \left[
\epsilon+(1-\epsilon)P\{d_{i,j}\le \bar d\}\right]
\]
depends on the particular specification of the distance $d_{ij}$.
Matching the link creation and removal processes, yields an
equation for the degree distribution $p(k)$. The probability
$P\{d_{i,j}\le \bar d\}$, in its turn, will itself depend on the
network density, i.e. on $p(k)$. Our approach will then have the
flavor of a self-consistent mean field approximation.

\section{Similarity by (chains of) acquaintances}
\label{sec:chain}

Consider first the simplest possible such specification where
$d_{ij}(t)$ is the (geodesic) distance between $i$ and $j$ on the
graph $g(t)$, neighbours $j$ of $i$ having $ d_{ij}(t)=1$,
neighbours of the neighbours of $i$ (which are not neighbours of $
i$) having $d_{ij}(t)=2,$ and so on.  If no path joins $i$ and $j$
we set $ d_{ij}(t)=\infty $.

This specific model describes a situation where the formation of new links
is strongly influenced by proximity on the graph. It is a simple manifestation
of our general idea that close interaction brings
about similarity -- here the two metrics coincide.
We set $\bar{d} \ge n-1$, the link formation process then discriminates between
agents belonging to the same network component (which are joined by at least
one path of links in $g$) and agents in different components. Distinct
components of the graph may, for example, represent different social groups.
Then Eq. (\ref{Pform}) captures the fact that belonging to the same social
group is important in the creation of new links (say, because it facilitates
control or reciprocity \cite{Coleman,An}).

\begin{figure}[tbp]
\includegraphics[width=0.9\columnwidth,clip]{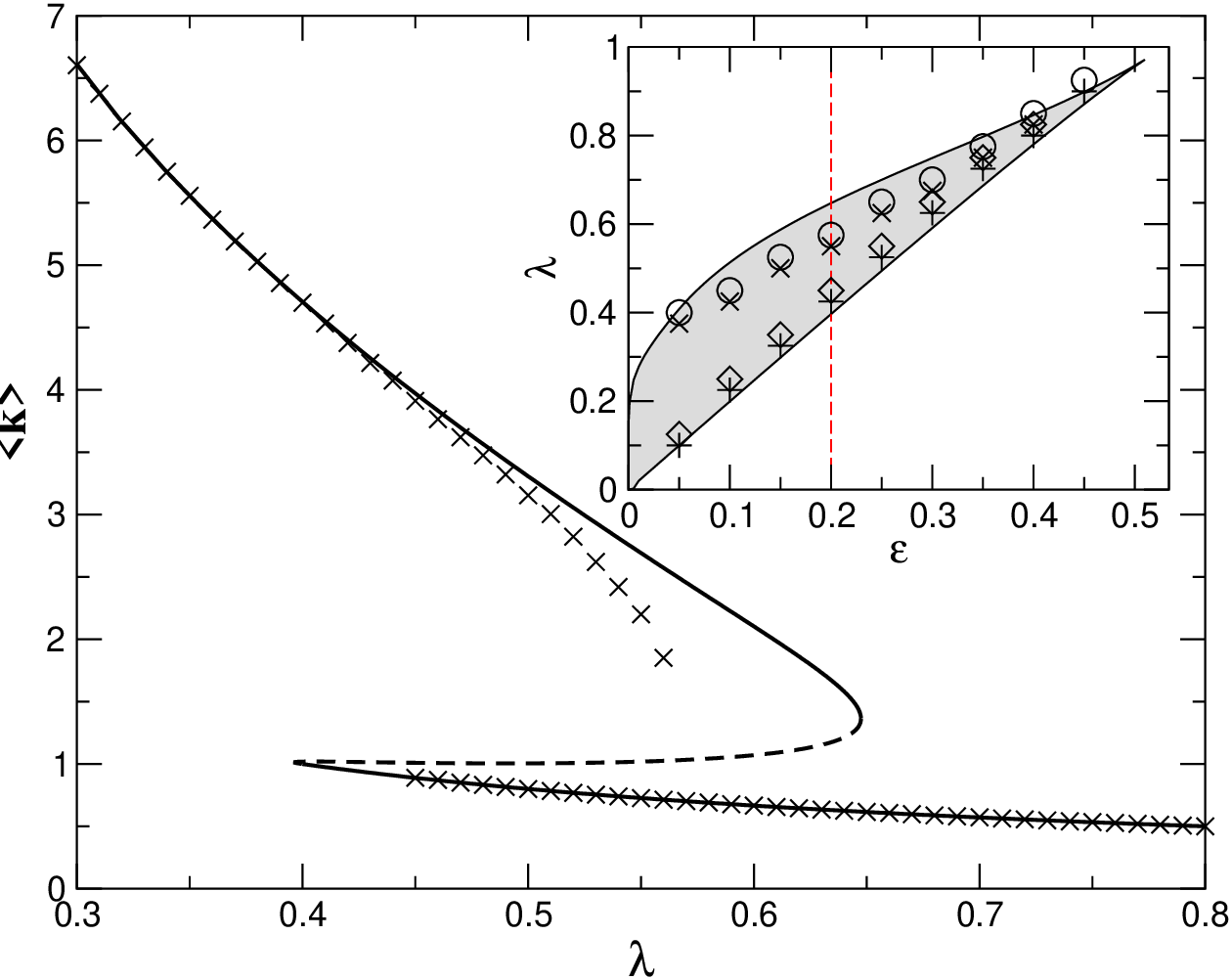}
\caption{ Mean degree $\left< k \right>$ as a function of
$\lambda$ ($\eta$ has been set to $1$) for $\epsilon =0.2$ when
$d_{ij}$ is the distance on the graph and $\bar{d} \ge n-1$. The
results of a mean field theory for $n=\infty $ (solid line) are
compared to numerical simulations ($\times$) starting from both
low and high connected states with $n=20000$. The dashed line
corresponds to an unstable solution of the mean field equations
which separates the basins of stability of the two solutions.  For
finite $n$ the low density state \textquotedblleft
flips\textquotedblright\ to the high density state when a random
fluctuation in $\left< k \right>$ brings the system across the
stability boundary (i.e. when a sizable giant component forms).
These fluctuations become more and more rare as $n$ increases.
 \emph{Inset}
: Phase diagram in mean field theory. Coexistence occurs in the
shaded region whereas below (above) only the dense (sparse)
network phase is stable. Numerical simulations (symbols) agree
qualitatively with the mean field prediction. The high (low)
density state is stable up (down) to the points marked with
$\times$ ($\diamond$) and is unstable at points marked with $\circ
$ ($+$). The behavior of $\left< k \right>$ along the dashed line
is reported in the main figure. } \label{fig_giant}
\end{figure}

Consider first what happens when $\eta / \lambda$ is small.  Let
$\left< k \right>$ be the average connectivity (number of links
per node) in the network.  The average rate $n \lambda \left< k
\right>/2$ of link removal is very high when $\left< k \right>$ is
significant. Consequently, we expect to have a very low $\left< k
\right>$, which in turn implies that the population should be
fragmented into many small groups. Under these circumstances, the
likelihood that an agent $i$ \textquotedblleft
meets\textquotedblright\ an agent $j$ in the same component is
negligible for large populations, and therefore new links are
created at a rate equal to $n \eta \epsilon $. By balancing link
creation and link destruction, the average number of neighbours of
an agent is $\left< k \right> = 2 \eta \epsilon / \lambda$, as is
indeed found in our simulations (Fig \ref{fig_giant}).

As $\eta / \lambda $ increases, the network density $\left< k
\right>$ increases gradually. Then, at a critical value $(\eta /
\lambda)_{1} = 1/2\epsilon$ -- when $\left< k \right> = 1$ -- a
giant component forms. The system makes a discontinuous jump (Fig.
\ref{fig_giant}) to a state containing a large and densely
interconnected community covering a finite fraction of the
population. If $\eta / \lambda$ decreases back again beyond the
transition point $(\eta / \lambda)_{1}$, the dense network remains
stable.  The dense network dissolves back into a sparsely
connected one only at a second point $(\eta / \lambda)_{2}$. This
phenomenology characterises a wide region of parameter space (see
inset of Fig. \ref{fig_giant}) and is qualitatively well
reproduced by a simple mean field approach.

It is worth mentioning that a similar
phenomenology occurs when $\bar{d}=2$, i.e. when links are
preferentially formed with \textquotedblleft friends of
friends \textquotedblright \cite{matteocomment}. In this case,
however, the probability that two arbitrary nodes $i$ and $j$ have
$d_{ij}=2$ is of order $1/n$ in a network with finite degree.
Hence for finite $\epsilon $ and $\lambda $ non-linear effect
manifest only for networks of finite sizes \cite{matteocomment}.

We finally mention that the model with $\bar d=2$ is reminiscent
of a model that was recently proposed \cite{PNAS} to describe a
situation where (as e.g. in job search \cite{Job}) agents find new
linking opportunities through current partners. In \cite{PNAS}
agents use their links to search for new connections, whereas here
existing links favour new link formation. In spite of this
conceptual difference, the model in Ref. \cite {PNAS} also
features the phenomenology (a)-(c) above, i.e. sharp transitions,
resilience, and phase coexistence.

\subsection{Mean Field Analysis}

The transition rate for the addition of a new link is
$w(k\rightarrow k+1)=2\eta \epsilon $ if the two agents are in
different components and $w(k\rightarrow k+1)=2\eta$ if they are
in the same component, where the factor $2$ comes because each
node can either initiate or receive a new link. In the large $n$
limit the latter case only occurs with some probability if the
graph has a giant component $\mathcal{G}$ which contains a finite
fraction $\gamma$ of nodes. For random graphs (see Ref.
\cite{randomGraph} for details) the fraction of nodes in
$\mathcal{G}$ is given by
\begin{equation}
\gamma=1-\phi (u)
\label{gammadef}
\end{equation}
where
\begin{equation}
\phi (s)=\sum_{k}p(k)s^{k}
\label{phidef}
\end{equation}
is the generating function
and $u$ is the probability that a link, followed in one direction, does not lead to the giant
component. The latter satisfies the equation
\begin{equation}
u=\phi ^{\prime }(u)/\phi^{\prime }(1).
\label{udef}
\end{equation}
Hence $u^{k}$ is the probability an agent with $k$ neighbours has no
links connecting him to the giant component, and hence is itself not part of the giant component.  Then the rate of
addition of links takes the form
\begin{equation}
w(k\rightarrow k+1)=2 \eta [\epsilon +(1-\epsilon )\gamma (1-u^{k})] .
\end{equation}
The stationary state condition of the master
equation leads to the following equation for $\phi (s)$
\begin{equation}
\lambda \phi ^{\prime }(s)=2 \eta [\epsilon +(1-\epsilon )\gamma ]\phi (s)-2 \eta (1-\epsilon )\gamma \phi (us)  \label{phis}
\end{equation}
which can be solved numerically to the desired accuracy. Notice that Eq.
(\ref{phis}) is a self-consistent problem, because the parameters $\gamma $
and $u$ depend on the solution $\phi (s)$. The solution of this equation is
summarised in Fig. \ref{fig_giant}. Either one or three solutions are found,
depending on the parameters. In the latter case, the intermediate solution
is unstable (dashed line in Fig. \ref{fig_giant}), and it separates the
basins of attraction of the two stable solutions within the present mean
field theory.

The solution is exact where there is no giant component, and
numerical simulations show that the mean field approach is very
accurate away from the phase transition from the connected to the
disconnected state. Near the transition to the disconnected state
our approximation, that an agent's degree fully specifies its
state, breaks down. This causes the theory to overestimate the
size of the coexistence region.

\section{Similarity of Knowledge/Technology levels}
\label{sect:h}

Next, we consider a setup where $d_{ij}$ reflects proximity of
nodes $i$, $j$ in terms of some continuous (non-negative) real
attributes, $h_{i}(t)$, $h_{j}(t)$. This case has been dealt with
in Ref. \cite{IJGT}, which provides a detailed socio-economic
motivation for the model. In short, the attribute $h_i$ could
represent the level of technical expertise of two firms involved
in an R\&D partnership, or the competence of two researchers
involved in a joint project. It could also be a measure of income
or wealth that bears on the quality and prospects of a bilateral
relationship.
We assume that each agent $i$ receives an attribute update (or
upgrade) possibility at a rate $\nu$. We focus on the case where
the dynamics of $h_i$ is much faster than that of the network
($\nu\gg\lambda,\eta$). In the opposite limit, links exits for too
short a time span to have any correlated effect on the dynamics of $h_i$. If
agent $i$ receives an update opportunity at time $t$, we posit
that

\begin{equation}
h_{i}(t^{+})=D\{h_{j},~j\in{\cal N}_i(t)\}+\eta _{i}(t)
\label{heq}
\end{equation}%
where $\eta _{i}(t)$ is a random term capturing the idiosyncratic
change of expertise due to $i$'s own (say research) efforts. In
Eq. (\ref{heq}) the function $D\{\cdot \}$ captures some process
of diffusion (e.g. sharing of knowledge) in the current
neighborhood $\mathcal{N}_{i}(t)=\{j:~ij\in g(t)\}$ of agent $i$.

We will take $\eta _{i}(t)$ to be Gaussian i.i.d. random variables
with zero mean and variance $\Delta$. This random idiosyncratic
term competes with the homogenising force of diffusion described
by the first term in Eq. (\ref{heq}). Concerning this term, we
will consider two alternative models.

\subsection{Best-Practise imitation}

The first one, which we will label Best-Practise Imitation (BPI)
has the revising player achieve a knowledge level equal to the
maximum available in her neighbourhood.
We have in mind a
situation where individuals aim at improving in the direction of
increasing $h_i$ and they may do this by some on-site
effort and also by learning from other individuals.

Formally, this is captured by the following definition:
\begin{equation}
D\{h_{j},~j\in{\cal N}_i(t)\cup\{i\}\} =\max_{j\in
\mathcal{N}_{i}(t)}h_{j}(t).  \label{deq}
\end{equation}%
Notice that if $i$ has no neighbor ($\mathcal{N}_{i}(t)=\emptyset$) then $D=h_i$.
This is equivalent to a directed polymer at zero temperature on the (dynamic) network $g(t)$ \cite{DPRM}.
A related model, using the idea of best practice imitation but with different noise has been studied in \cite{majumdar}, but for randomly chosen neighbours at each interaction (i.e. no network).

\begin{figure}[tbp]
\includegraphics[width=0.9\columnwidth,clip]{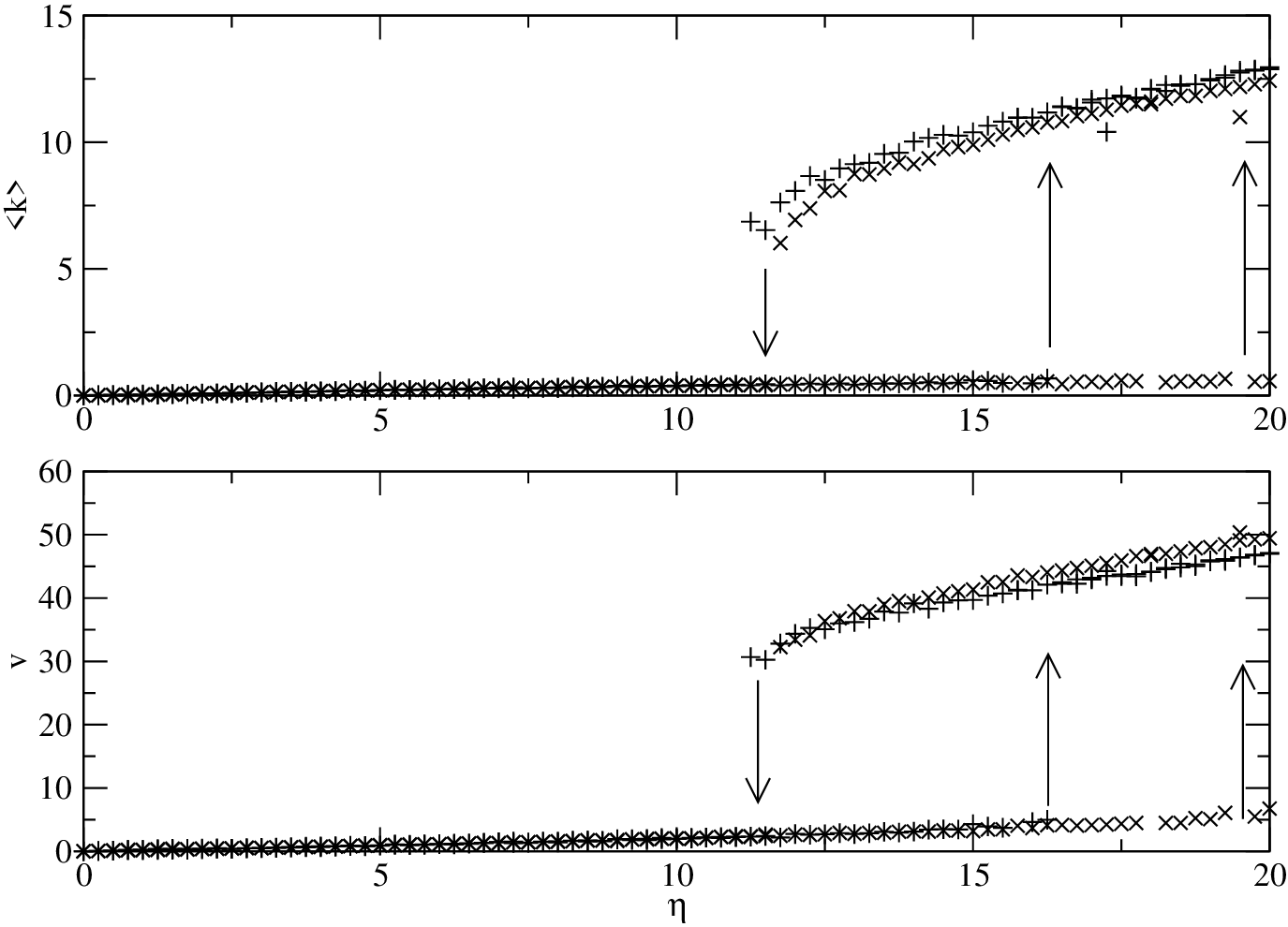}
\caption{
Mean degree $\left< k \right>$ (top) and growth rate $v$ (bottom) as a function of $\eta$, found from numerical simulations of the model with Eq. (\ref{deq}).
Shown are simulations with $n=500$ (plusses) and $1000$ (crosses).
Arrows denote the approximate point at which the system jumps from one phase to the other (this point can be dependent on $n$).  Here $\epsilon=0.001$, noise strength $\Delta =0.1$, similarity threshold $\bar d =2$.
The system was run up to $t = 1000$ for equilibration, then from $t = 1000$ to $t=1100$ for data taking.
}
\label{fighKPZ}
\end{figure}

Fig. \ref{fighKPZ} reports typical results of simulations of this model.
As in the two previous models, we find a discontinuous transition between a sparse and a dense network state, characterised by hysteresis effects.
When the network is sparse, diffusion is ineffective in homogenising growth.
Hence the distance $d_{ij}$ is typically beyond the threshold $\bar{d}$, thus the link formation process is slow.
On the other hand, with a dense network, diffusion keeps the gaps between the $h_{i}$s of different nodes small, which in turn has a positive effect on network formation.
As before, the phase transition and hysteresis is a result of the positive feedback that exists between the dynamics of the $h_{i}$ and the adjustment of the network.
In the stationary state we find that $h(t)\equiv \left\langle h_{i}(t)\right\rangle $ grows linearly in time, i.e. $h_i(t)\simeq v t$.
Notably, the growth process is much faster (i.e. $v$ is much higher) in the dense network equilibrium than in the sparse one, as shown in the lower panel of Fig. \ref{fighKPZ}.

This model exhibits an interplay between the process on the
network - the $h_i(t)$ which depends on the network - and the
network evolution which depends on the $h_i(t)$.  It is this
interdependence and the corresponding positive feedback which
produces the discontinuous transition and phase coexistence.

The similarity of the behavior of this model with that of the
previous section can be understood by analyzing a particular
limit. Consider indeed the case where $\eta_i=1$ with probability
$a$ and $\eta_i=0$ otherwise. When $\nu a\ll\eta$ innovations take
place at a rate much smaller that that over which new links form.
In the limit where the dynamics of $h_i$ is fast enough
($\nu\gg\eta$), we can assume that each new innovation (i.e. each
event $\eta_i=1$) taking place on a connected component
instantaneously propagates to the entire set of connected nodes.
Hence, if $\bar d< 1$, link creation will occur with probability
one on nodes in the same connected component, whereas nodes in
different components will likely have distinct values of $h_i$, so
that links will form at rate $\eta\epsilon$. Note also that, in
this particular limit, the growth rate $v$ is proportional to the
size of the largest connected component.

\subsection{Conforming to neighbors}

The second alternative considered has diffusion
embody a uniform merging of the neighbourhood's levels, formalised as follows%
\begin{equation}
D\{h_{j}
\} =\left\{\begin{array}{cc}
  \frac{1}{|\mathcal{N}_{i}|}\sum_{j\in \mathcal{N}_{i}(t)}h_{j}(t) &
  \mathcal{N}_{i}(t)\not =\emptyset \\
  h_i & \mathcal{N}_{i}(t) =\emptyset
\end{array}\right.
\label{udeq}
\end{equation}%
where $|\mathcal{N}_{i}(t)|$ is the number of agents in $i$'s
neighborhood. This second formulation can be conceived as
reflecting a process of opinion exchange (with no idea of relative
\textquotedblleft advance\textquotedblright\ in the levels
displayed by different individuals) \cite{weisbuch,demarzo}.  Alternatively, it could
be viewed as reflecting a context where interaction payoffs are
enhanced by compatibility (say, of a technological nature) and
agents will naturally tend to adjust towards their neighbours'
levels. In these cases, interaction promotes conformity and
conformism constraints the creation of new links. At any rate,
this specification of the model allows us to understand how 
the results of the previous section depend on the directionality
of the diffusion process.

\begin{figure}[tbp]
\includegraphics[width=0.9\columnwidth,clip]{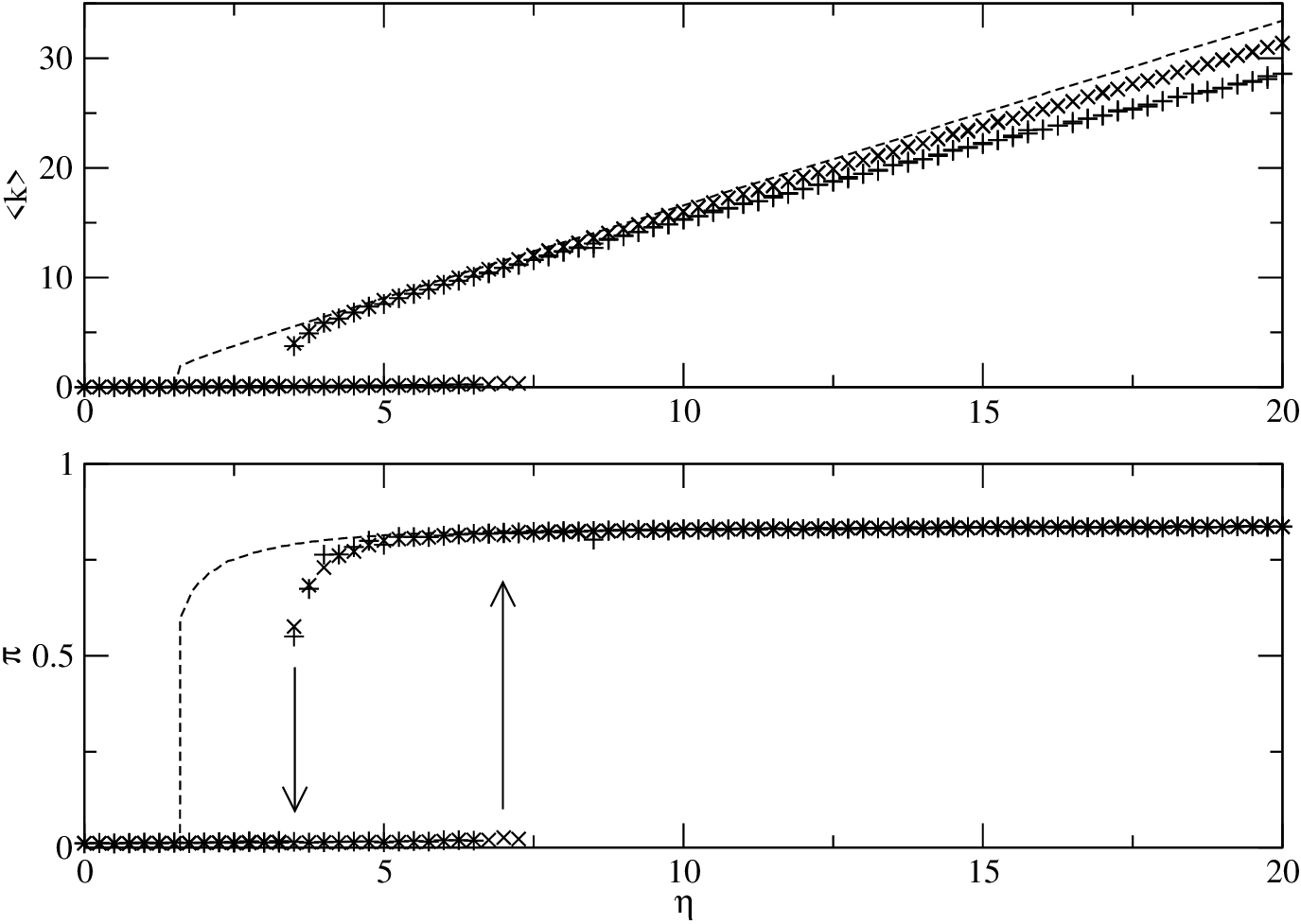}
\caption{
Mean degree $\left< k \right>$ (top) and probability that two randomly chosen nodes are within $\bar d$ of each other, $\pi$ (bottom), as a function of $\eta$.
Shown are simulations with $n=200$ (plusses) and $n=500$ (crosses).
Also theory for the high connected phase (dashed line).
For large $\eta$, the data points converge towards the theory curve as $n$ increases.
Arrows denote the approximate point at which the system jumps from one phase to the other.
Here $\epsilon=0.001$, noise strength $\Delta =1$, similarity threshold $\bar d =2$.
The system was run up to $t = 1000$ for equilibration, then from $t = 1000$ to $t=1100$ for data taking.
}
\label{fighEW}
\end{figure}

Fig. \ref{fighEW} shows that this model exhibits the same generic
phenomenology of a sharp transition and the coexistence of sparse
and dense network phases. The key consideration, in this case, is
that when the link density is high, the distribution of $h_i$ in
the population is narrow and hence link creation proceeds at a
relatively fast rate.

This intuition is captured by a simple mean field approach. We
will assume that the network can be well approximated by an
Erdos-Renyi random graph with average degree $\avg{k}$. When
$\nu\gg \eta,\lambda$, we can assume that the distribution of
$h_i$ adjusts adiabatically to the changing network. In this limit
the dynamics is well described by the Edwards-Wilkinson Langevin
equation \cite{footnote3}.

\begin{equation}
\dot h_i=-\frac{\nu}{|{\cal
    N}_i|} \sum_{j\in{\cal N}_i} (h_j-h_i) +\zeta_i \equiv -\nu\sum_j
{\cal L}_{i,j}h_j+\zeta_i \label{langevin}
\end{equation}
This can be seen by considering a small time interval $dt$. If
$\nu dt\gg 1$, the number of updates on each site is large and can
be approximated with the central limit theorem with the two terms
in Eq. (\ref{langevin}). In this equation, $\zeta_i(t)$ is a white
noise term with zero average and
$\avg{\zeta_i(t)\zeta_j(t')}=\nu^2\Delta\delta_{i,j}\delta(t-t')$
and we have introduced the (normalized) Laplacian matrix of the
graph ${\cal L}$. The dynamics of this model is easily integrated
in the normal modes of the diffusion operator. In other words, let
$\vec v^\mu$ be the eigenvectors of ${\cal L}$, i.e. ${\cal L}\vec
v^\mu = \mu \vec v^\mu$, then the normal modes $h^\mu=\sum_i
v_i^\mu h_i$ satisfy

\begin{equation}
\dot h^\mu =- \nu\mu h^\mu-\zeta^\mu
\end{equation}
where, in view of the orthogonality of the transformation
$i\to\mu$, $\zeta^\mu$ is again a white noise with the same
statistical properties of $\zeta_i$. The fluctuations of $h^\mu$
in the stationary state are $\avg{(h^\mu-\avg{h^\mu})^2}=\frac{\nu
\Delta}{2\mu}$. Back transforming to the variables $h_i$ one finds
that

\begin{equation}
\avg{(h_i-\avg{h_i})^2}=\sum_{\mu>0} \frac{\nu
  \Delta}{2\mu} =\frac{\nu \Delta}{2}\int
\frac{d\mu}{\mu}\rho(\mu)
\end{equation}

\noindent where $\rho(\mu)$ is the density of eigenvalues of the
Laplacian matrix, which has been computed in the limit
$n\to\infty$ \cite{dgms_spectra}. Notice that we disregard finite
size clusters, which contribute to a $\mu=0$ peak in the spectrum,
assuming that the $h_i$ value of these nodes is broadly
distributed so that $d_{i,j}>\bar d$ whenever $i$ or $j$ are not
in the giant component. There is no simple closed form for
$\rho(\mu)$, so one has to resort to numerical calculation. To our
level of approximation, it is sufficient to stick to a simple
approximation \cite{dgms_spectra}, where
\begin{equation}
\rho( \mu ) = - { 1 \over \pi } \text{ Im } {1 \over \mu - T(\mu)
} . \label{rho}
\end{equation}
and $T(\mu)$ is the solution of
\begin{equation}
T(\mu) = {1 \over \avg{k}} \sum_k { k P(k) \over k \mu +i \epsilon
-(k-1) T(\mu) } \label{Tlam}
\end{equation}
with $\epsilon \to 0^+$. The key features are that:
\begin{itemize}
  \item the integral
\[
R(\avg{k})=\int \frac{d\mu}{\mu}\rho(\mu)
\]
for Erdos-Renyi graphs, is a function of the average degree
$\avg{k}$ alone.
  \item The function $R(c)$ decreases monotonically and it diverges as
  $c\to 1^+$, when the giant component vanishes
\end{itemize}

This allows us to estimate the probability
\begin{equation}
P\{|h_i-h_j|<\bar d\}=\theta(\avg{k}-1){\rm erf}\left[\bar
d/\sqrt{2\nu \Delta R(\avg{k})}\right]
\end{equation}
where the $\theta$ function implies that this probability vanishes
for disconnected graphs. Equating the link formation and removal
rate, finally provides an equation for $\avg{k}$ which reads
\begin{equation}\label{balance}
  \frac{\lambda}{2\eta}\avg{k}=\epsilon+(1-\epsilon)
  \theta(\avg{k}-1){\rm erf}\left[\bar
d/\sqrt{2\nu \Delta R(\avg{k})}\right].
\end{equation}
Fig. 3 reports the numerical solution of this equation for the
same parameters as in the simulations. This agreement is
reasonably good in view of the approximations made. Again the mean
field approach overestimates the size of the coexistence region.
The mean-field calculation reproduces the main qualitative
behavior, even though it (again) overestimates the size of the
coexistence region.

The emergence of features {\em (a)-(c)} depends crucially on the
divergence of $R(\avg{k})$ on the average degree when
$\avg{k}\approx 1$. This divergence gets smoothed when $\nu$
decreases, which suggests that the discontinuous transition should
turn into a smooth crossover beyond a critical value $\nu_c$. This
scenario, which is reminiscent of the behavior at the liquid-gas
phase transition, is indeed confirmed by numerical simulations.

\section{Coordinating in a changing world}
\label{sec:potts}

We now consider a further specialisation of the general framework
where link formation requires some form of coordination,
synchronisation, or compatibility.  For example, a profitable
interaction may fail to occur if the two parties do not agree on
where and when to meet, or if they do not speak the same
languages, and/or adopt compatible technologies and standards. In
addition, it may well be that shared social norms and codes
enhance trust and thus are largely needed for fruitful
interaction.

To account for these considerations, we endow each agent with an
attribute $x_{i}$ which may take one of $q$ different values,
$x_{i}\in \{1,2,\ldots ,q\}$.  $x_{i}$ describes the internal
state of the agent, specifying e.g. its technological standard,
language, or the social norms she adopts.  The formation of a new
link $ij$ requires that $i$ and $j$ display the same attribute,
i.e. $x_{i}=x_{j}$. This is a particularisation of the general Eq.
(\ref{Pform}) with $d_{ij}=1-\delta_{x_{i},x_{j}}$ and
$0<\bar{d}<1$. For simplicity we set $\epsilon =0$ since in the
present formulation there is always a finite probability that two
nodes display the same attribute and hence can link.  We assume
each agent revises its attribute at rate $\nu$, choosing $x_i$
dependent on its neighbours' $x_j$s according to:
\begin{equation}
P\{x_{i}(t)=x\}=\frac{1}{Z} \exp \left[ \beta \sum_{j:ij\in g(t)}\delta _{x,x_{j}(t)}
\right]   \label{potts}
\end{equation}
where $\beta $ tunes the tendency of agents to conform with their
neighbours and $Z$ provides the normalisation. This adjustment
rule coincides with the Kawasaki dynamics, which is known to
sample the equilibrium distribution of the Potts model of
statistical physics \cite{Baxter} with temperature $T=1/( k_B \beta )$. Eq.
(\ref{potts}) has been used extensively, mainly for $q=2$, in the
socio-economic literature as a discrete choice model
\cite{Blume,Durlauf,Young}.

This model describes a situation where agents are engaged in
bilateral interactions which however require a degree of
coordination among partners (i.e. $x_i=x_j$). The agents attempt
to improve their situation both by coordinating their value of
$x_i$ with that of neighbours and by searching for neighbours in
their same state and linking with them. Link removal models decay
of links, e.g. due to obsolescence considerations. The stochastic
nature of the choice rule (\ref{potts}) captures a degree of
volatility or un-modelled features on which the interaction
depends (e.g. one might think that agent $i$ might have some
advantage for choosing a given value of $x_i$ at a particular
time). From the point of view of statistical physics, the presence
of a non-zero ``temperature'' prevents the system from getting
stuck in imperfect states. We will consider these effects in more
detail below (see section \ref{tzeroexactsolnsection} and Fig.
\ref{freezingfig}) when discussing the case $\beta \to \infty$ in
greater detail.

This is another manifestation of our general idea that
network-mediated contact favors inter-node similarity. As in
Section \ref{sect:h}, we focus on the case where such a
similarity-enhancing dynamics proceeds at a much faster rate than
the network dynamics. That is, $\nu \gg \eta,\lambda$ so that, at
any given $t$ where the network $g(t)$ is about to change, the
attribute dynamics on the $x_{i}$ have relaxed to a stationary
state. The statistics of this state are those of the Potts model
on the graph $g(t)$. For random graphs of specified degree
distribution $p(k)$, the necessary statistics of the Potts model
can be found \cite{DGM,EM} and this makes an analytic approach to
this model possible. We shall first discuss an approximate theory
to the general case and then focus on a particular limit where the
model can be solved exactly.

\subsection{Method of Solution}

\begin{figure}[tbp]
\includegraphics[width=0.9\columnwidth,clip]{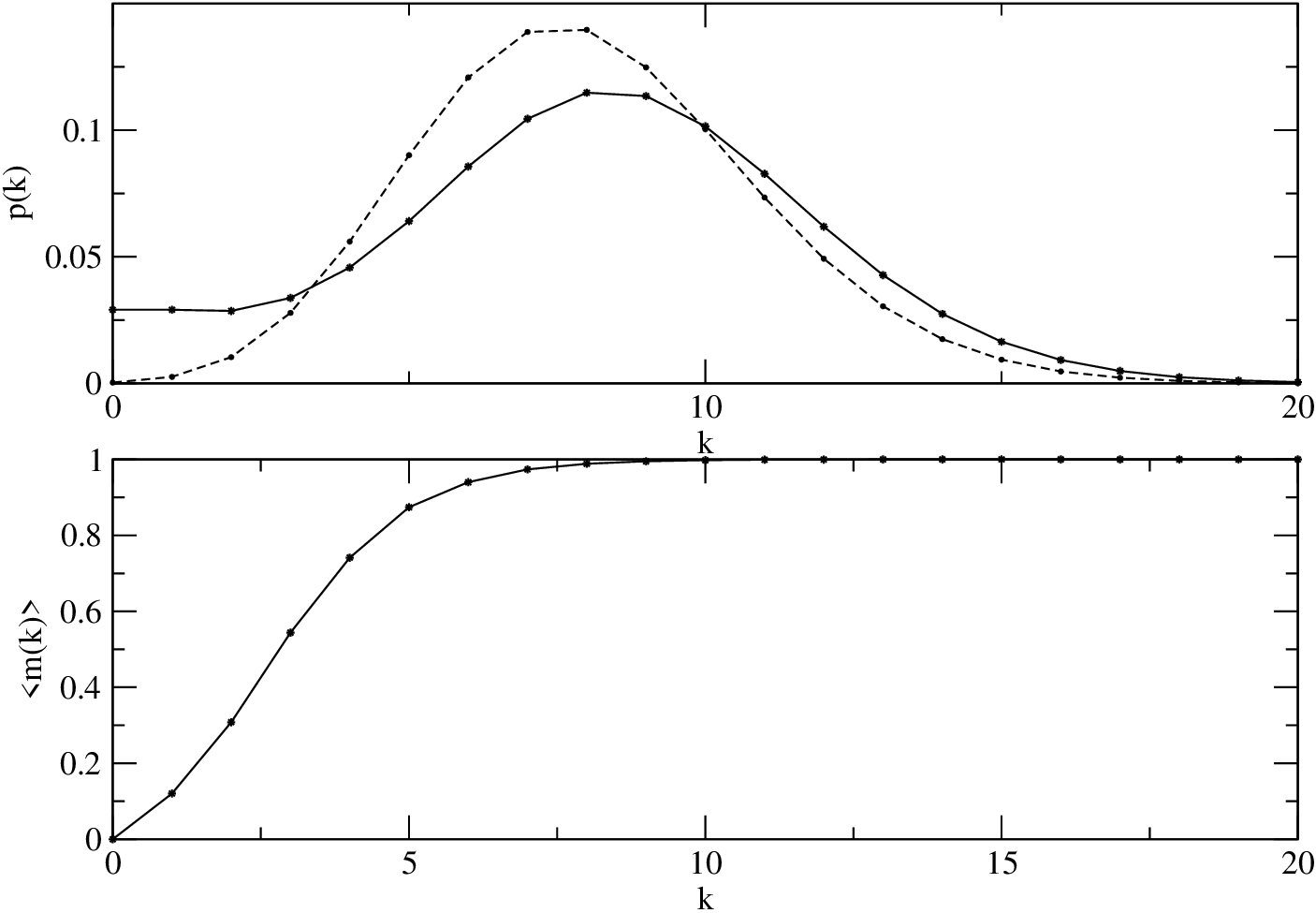}
\caption{  Results from the analytic solution.  Upper plot: Plot
of the degree distribution, $p(k)$ for $\eta/\lambda=4$, showing
also the Poisson distribution with the same $\left< k \right>$ for
comparison (dashed lines).  Note that the degree distribution is
{\it not} Poissonian. Lower plot: Plot of the average
magnetisation of a node of degree $k$ for $\eta/\lambda=4$),
showing that more highly connected nodes are, on average, more
magnetised. $q=10$. 
 }
\label{pofkANDmofkTMP2fig}
\end{figure}

Again we rely on the random graph approximation where the network
is completely specified by the degree distribution $p(k)$. Now
however the probability of two nodes being in the same state ${\it
if}$ they are both in the giant component depends on the
magnetisation of the giant component.  The master equation for a
general node of degree $k$ is,
\begin{eqnarray}
\dot p(k) =  &&\lambda (k+1) p(k+1)   +2 \eta p(k-1) \pi(k-1) \nonumber \\
             &&-\lambda k p(k)         -2 \eta p(k) \pi(k)
\label{pottsdyn}
\end{eqnarray}
where $\pi(k)$ is the probability that a node of degree $k$ is in
the same state as a randomly chosen node. This crucially depends
on whether the Potts spins are ordered or not. Indeed, for
sufficiently high $\beta$ the equivalence between the different
$q$ spin states is broken in the stationary state of the Potts model
with temperature $1/( k_B \beta )$. This is signalled by a non-zero value
of the magnetization
\begin{equation}\label{pottsm}
  m=\frac{q\langle\delta_{x,1}\rangle -1}{q-1}
\end{equation}
where the average is both on the nodes of the giant component and
on the stationary distribution. Without loss of generality, we can
assume that $x=1$ is the state which is selected and $m>0$ implies
$\langle\delta_{x,1}\rangle>1/q$. Then it is easy to see that
\begin{equation}
\pi(k) = {1 \over q} +{ q-1 \over q }\gamma (1-u^k) m(k) m
\label{pikpotts}
\end{equation}
where $\gamma$, $u$ are defined in Eqs.
(\ref{gammadef},\ref{udef}) and
$m(k)=(q\langle\delta_{x,1}|k\rangle -1)/(q-1)$. Note that
$\pi(k)$ and $m(k)$ depend on $k$, more highly connected nodes
being on average more coordinated/magnetised (see figure
\ref{pofkANDmofkTMP2fig}).

With these equations we can find $p(k)$ and $\pi(k)$ iteratively.
Starting from a given $p(k)$, we first compute the properties of
the Potts model on a random network with such a degree
distribution, from which we get $\pi (k)$ in Eq. (\ref{pikpotts}).
This with Eq. (\ref{pottsdyn}) in the stationary state ($\dot
p(k)=0$) allows us to estimate $x_k=p(k)/p(0)$ from the equation
\[
x_{k+1}=\frac{[2 \eta \pi(k) +\lambda k] x_k - 2
  \eta \pi( k- \! 1)x_{k-1}}{\lambda (k \! + \! 1)}
\]
iteratively, in terms of $x_0=1$ and $x_1=2\eta\pi(0)/\lambda$.
Normalization, yields a new estimate of the degree distribution
$p(k)=x_k/\sum_h x_h$. We repeat this cycle until a stable
solution $p(k)$ is found.

\begin{figure}[tbp]
\includegraphics[width=0.9\columnwidth,clip]{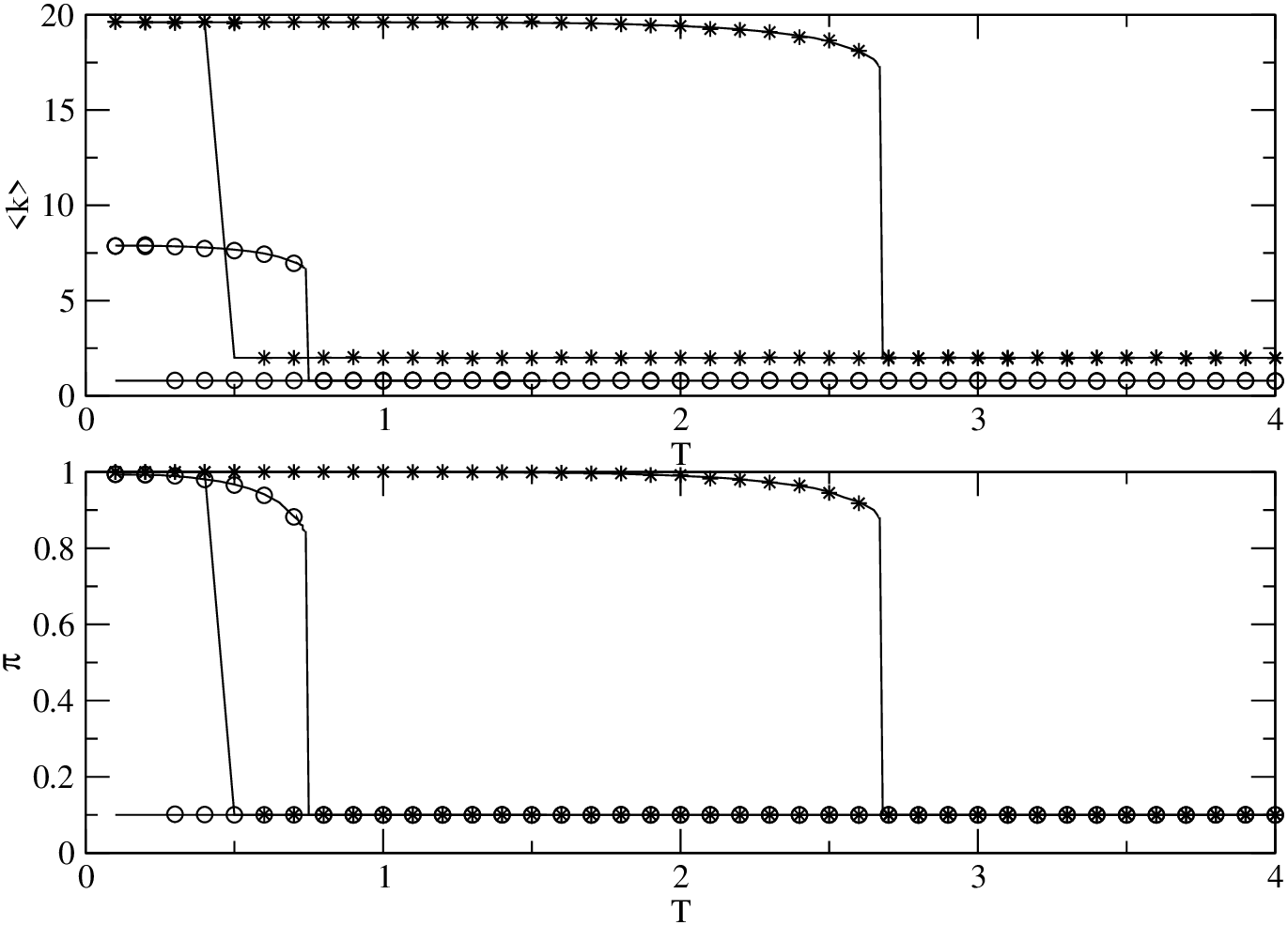}
\caption{ Upper plot:  Plots of average degree ($\left< k
\right>$) as a function of 'temperature' $T=1/\beta$. Lower plot:
Plots of the probability that two randomly chosen nodes are in the
same state ($\pi$), as a function $1/\beta$. All plots are for:
$\eta/\lambda=4$ (lower curves) and $\eta/\lambda=10$ (higher
curves).  Points are results of simulations.  $q=10$.
for $n=1000$. } \label{cvstemp_etais4and10_nis1000fig}
\end{figure}

Figure \ref{cvstemp_etais4and10_nis1000fig} shows $\left< k
\right>$ and $\pi$ plotted against temperature for both
simulations and theory.  Note that the agreement is excellent
despite the approximation made. As before, for high $\eta /
\lambda$ there is a highly connected network with a giant
component, for low $\eta / \lambda$ the network is sparsely
connected.  For intermediate values of $\eta / \lambda$ the two
states coexist and which one is found depends on the initial
conditions for $p(k)$.

\begin{figure}[tbp]
\includegraphics[width=0.9\columnwidth,clip]{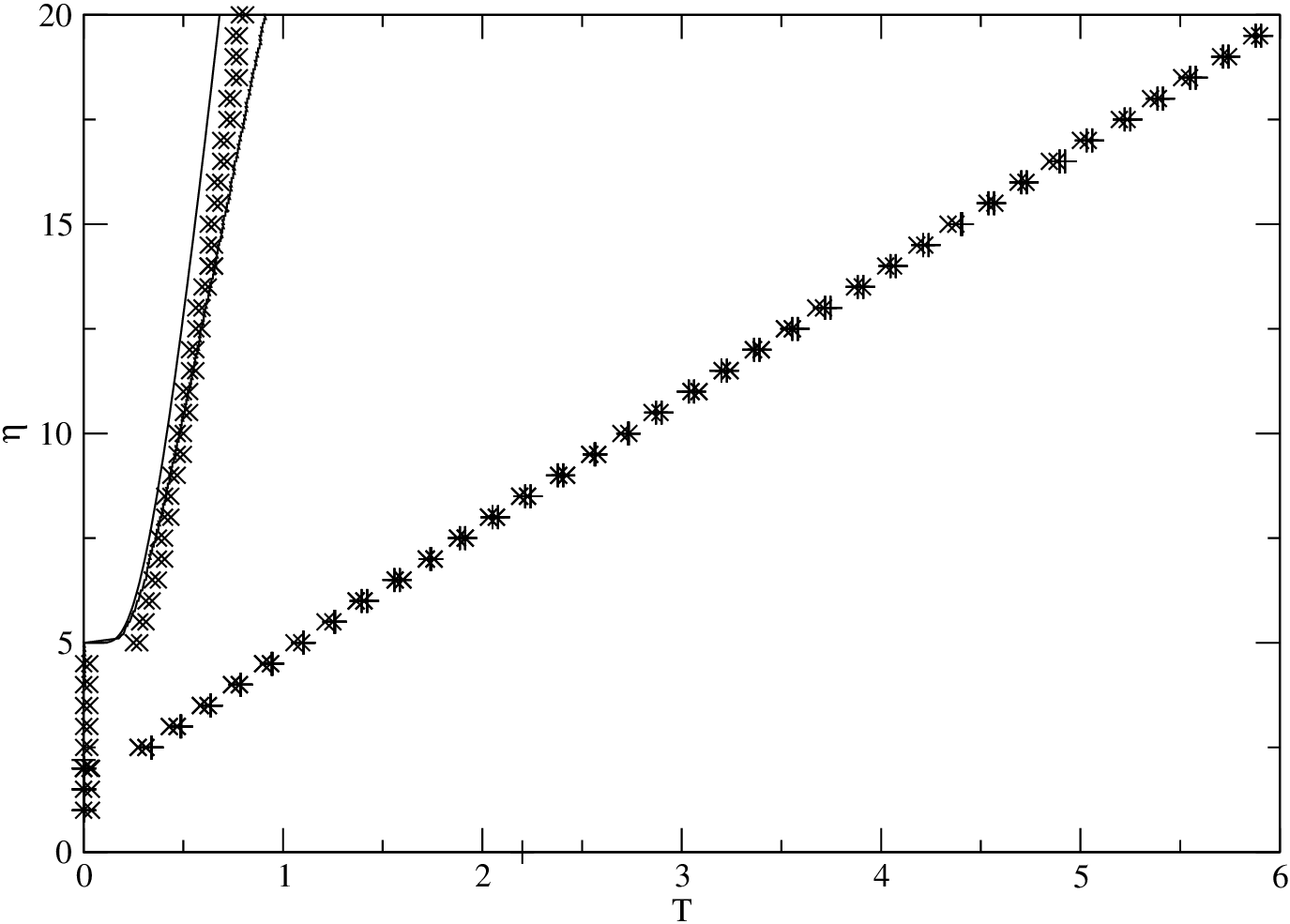}
\caption{
Phase diagram in $\eta$ ($\lambda=1$) and $T$, showing the high and low connectivity phases and the hysteretic region.
Crosses are simulations ($n=5000, 10000$), plusses are theory, the points are in pairs, one on each side of the phase line.
The curves are theory, the leftmost curve is $\eta_c=(q/2) (\exp(1/T)+q-1) / (\exp(1/T)-1)$ which is the expected transition line if the system gets stuck in the metastable, unmagnetised state.
The right curve is found using the normal method described above. \\
Upper left: higher connectivity, coordinated region.\\
Lower right: lower connectivity, uncoordinated region.\\
The central region is the hysteretic region.
}
\label{pottsongrowingnetworkphasefig}
\end{figure}

Figure \ref{pottsongrowingnetworkphasefig} shows a phase diagram
(simulations and theory) for $\beta$ and $\eta / \lambda$.  It can
be seen that the theory is rather close to the simulation results.
The low uncoordinated region is a Poisson random graph with
$\left< k \right> = 2 \eta /(\lambda q)$. Starting in this state,
the transition to the highly connected, magnetised state, can only
occur when there is a giant component, i.e. $\left< k \right>
> 1$ so $\eta > \lambda q/2$.  Hence the lowest point of the high
connectivity region of figure \ref{pottsongrowingnetworkphasefig}
is at $\eta = 5$ and $T=0$. Although $\nu \gg \eta,\lambda$, the
simulations of the Potts model can still get into a metastable
unmagnetised state, even below the transition temperature
\cite{EM}.  The temperature $T$ at which this metastable state
becomes unstable is given by \cite{EM}
\begin{equation}
\exp(1/T) = {\left< k^2 \right> +(q-2)\left< k \right> \over \left< k^2 \right> -2\left< k \right>}.
\label{}
\end{equation}
The (uncoordinated) graph is Poissonian, $\left< k^2 \right>
= \left< k \right>^2 +\left< k \right>$.  Thus the transition
curve is
\begin{equation}
\eta_c = {q \over 2} {\exp(1/T)+q-1 \over \exp(1/T)-1}.
\label{}
\end{equation}
Monte-Carlo simulations show that this theoretical line is slowly approached
as $n$ is increased.

\section{Exact Solution for $T=0$. \label{tzeroexactsolnsection}}
\label{sec:exact}

The model with $\beta\to\infty$ and $\epsilon=0$ can be described
exactly. 
We assume that in the initial state, links exist {\it only} between nodes that have the same spin, $\sigma_i = sigma _j$.
The key intuition is that the spin of site $i$ can change
only if $k_i=0$, i.e. if the site is isolated. Hence we can
classify sites in disjoint subsets $N=N_0\bigcup_{\sigma=1}^q
N_\sigma$ where
\begin{eqnarray}\label{Ni}
N_0 &&=\{i:k_i=0\} \nonumber \\
N_\sigma &&=\{i:k_i>0,\sigma_i=\sigma\} ~~~~~ \sigma =1,\ldots,q
\end{eqnarray}
The spins $\sigma_i$ are frozen for all nodes $i\in N_\sigma$ with
$\sigma>0$ whereas nodes in $N_0$ have spin which are randomly
updated at a fast rate. Because $\epsilon=0$, links can only be
formed between nodes $i$ and $j$ which are either both in the same
component $N_\sigma$ with $\sigma>0$, or both in $N_0$ provided
they have the same spin, or if one is in $N_\sigma$ and one is in
$N_0$, but has spin $\sigma_j=\sigma$. No link can be formed
between $i\in N_\sigma$ and $j\in N_{\sigma'}$ with
$\sigma,\sigma'>0$.

When a link with a node in $N_0$ is formed, one or two nodes pass
from $N_0$ to some $N_\sigma$. Likewise nodes of $N_\sigma$ which
lose their links move to $N_0$. Such a dynamics, in the limit
$n\to\infty$ is captured by the following evolution for the
fraction $n_\sigma$ of nodes in set $N_\sigma$ ($\sigma\ge 0$)

\begin{equation}\label{dyn}
\dot n_\sigma =  {\frac{2 \eta}{q}}n_0n_\sigma +{\frac{2
\eta}{q^2}} n_0^2-\lambda p_{\sigma,1} n_\sigma ,~~~~\sigma\ge 1.
\end{equation}

\noindent Here $p_{\sigma,k}$ is the degree distribution of nodes
in $N_\sigma$, and
\begin{equation}\label{n0}
  n_0=1-\sum_{\sigma=1}^q n_\sigma
\end{equation}
is fixed by the normalization. The first term in Eq. (\ref{dyn})
arises from the process where a node of degree zero joins a node
of type $\sigma$. The factor 2 is present because either node
might have initiated the link. The second term is the process
where a node of degree zero joins another node of degree zero. The
factor 2 is present because $n_{\sigma}$ increases by 2. The
dynamics of the degree inside a component is just that leading to
a random Poissonian graph for $k_i>0$. Hence $p_{\sigma,k}$ is
given by
\begin{equation}\label{pks}
p_{\sigma,k}=\frac{c_\sigma^k}{(e^{c_\sigma}-1)k!},~~~~~k\ge 1
\end{equation}
where the average degree
\begin{equation}\label{csig}
  c_\sigma=\frac{2 \eta n_0}{\lambda q} +\frac{2 \eta n_\sigma}{\lambda }
\end{equation}
is obtained by balancing the average link creation rate
$2\eta(n_\sigma+n_0/q)$ with the link removal rate $\lambda
c_\sigma$, inside the component $N_\sigma$. The equations above
allow we to recast the dynamics in the form

\begin{equation}\label{nsigdyn1}
\dot n_\sigma =  { \lambda^2 c_{\sigma} \over
2\eta [1-\exp(-c_{\sigma})]} \left[ {2\eta n_0 \over \lambda q}
-c_{\sigma} \exp(-c_{\sigma}) \right].
\end{equation}
From this it is clear that in the stationary states $c_{\sigma}$
obeys:

\begin{equation}\label{bob1}
c_{\sigma} \exp(-c_{\sigma})  = {2 \eta n_0 \over \lambda q}
\end{equation}
for $\sigma=1$ to $q$. Moreover Eq. (\ref{csig}) implies with the
constraint
\begin{equation}\label{bob2}
\sum_{\sigma=1}^q c_{\sigma}  = {2 \eta \over \lambda}.
\end{equation}

In order to build a solution, let us notice that Eq. (\ref{bob1})
has two solutions (provided that $2\eta n_0\le \lambda q e^{-1}$)
which we denote $c_+ > 1$ and $c_- < 1$. Hence solutions can be
specified in terms of the number $\ell$ of components with
$c_\sigma=c_+$. Then Eq. (\ref{bob2}) becomes
\begin{equation}\label{bob3}
\ell c_+  +(q-\ell) c_- = {2 \eta \over \lambda}
\end{equation}
Equations (\ref{bob1}, \ref{bob3}) can be solved for any value of
$\eta / \lambda$ and $\ell$.

The degree distribution for the whole network is given by:
\begin{equation}\label{}
p(k) = {n_0 \over q} \left[ \ell { {c_+}^k \over k! } + (q-\ell) {
{c_-}^k \over k! }   \right].
\end{equation}
The average degree can be written as
\begin{equation}\label{}
\left< k \right>  = {\ell c_+^2 + (q-\ell) c_-^2 \over \ell c_+ +
(q-\ell) c_-}
\end{equation}
We will now show that only the solutions with $\ell=0$ and $1$ are
dynamically stable. These are those which describe the behavior of
the model.

\subsection{Stability Analysis}

The dynamics can be written as $\dot n_\sigma =
f(n_0,n_{\sigma})$. Let $n_{\sigma} = {\bar n}_{\sigma}
+\epsilon_{\sigma}$ where $\bar n_{\sigma}$ is the solution
derived above (i.e. $f(\bar n_\sigma,\bar n_0)=0$) with
$c_\sigma=c_+$ for $\sigma\le \ell$ and $c_\sigma=c_-$ for
$\sigma>\ell$. Here $\epsilon_\sigma$ is a small perturbation
which, to leading order, satisfies

\begin{equation}\label{perturb}
\dot \epsilon_\sigma = { \lambda } \sum_{\nu=1}^q{T_{\sigma,\nu}}
\epsilon_\nu
\end{equation}
where ${\bf T}$ has matrix elements
\begin{equation}\label{Tij}
T_{\sigma,\nu} =  { c_{\sigma}( c_\sigma -1 ) \over
\exp(c_{\sigma})-1 }\delta_{\sigma,\nu}-\frac{1}{q} \left[
\frac{c_\sigma^2}{\exp(c_{\sigma})-1}+c_\sigma\right]
\end{equation}
A solution is stable if all eigenvalues of ${\bf T}$ are negative.
For the $\ell=0$ solution, $c_\sigma=c_-=2\eta/(\lambda q)$ for
all $\sigma$, we find $q-1$ ``transverse'' eigenmodes
($\sum_\sigma\epsilon_\sigma =0$)  with eigenvalue
$\mu=-(1-c_-)/(e^{c_-}-1)$ and one ``longitudinal'' mode
($\epsilon_\sigma=\epsilon$) with $\mu=-c_-/(1-e^{-c_-})$. Both
are stable ($\mu<0$) so the $\ell=0$ solution is always stable, as
long as $c_-<1$, i.e. for $2\eta<\lambda q$ (see Eq. \ref{bob3}).

It is also easy to find an unstable mode for solutions with
$\ell\ge 2$ components in the $c_+$ state. Let $c_\sigma=c_+$ for
$\sigma\le \ell$ and $c_\sigma=c_-$ otherwise and consider
``transverse'' perturbations with $\epsilon_\sigma=0$ for
$\sigma>\ell$ and $\sum_\sigma \epsilon_\sigma=0$. These describe
density fluctuations among $c_+$ components. We find
$\dot\epsilon_\sigma=\lambda\mu\epsilon$ with $\mu=
c_+(c_+-1)/(e^{c_+}-1)>0$. This means that any perturbation of
$\ell>1$ solutions with an imbalance between two or more
components with $c_\sigma=c_+$ will grow exponentially, thus
leading to the collapse of all but one of the components.

\subsection{The $\ell=1$ solution}

Combining equations (\ref{bob1}, \ref{bob3}), the equation for
$c_-$ with $\ell =1$ can be written as
\begin{equation}\label{ell1sol}
c_- -\frac{2\eta/\lambda-qc_-}{e^{2\eta/\lambda-qc_-}-1}=0.
\end{equation}

This equation has no solution for $\eta<\eta_c$, where $\eta_c$ is
the point where the maximum of the l.h.s. of Eq. (\ref{ell1sol})
as a function of $c_-$, becomes zero.  Beyond this point
($\eta>\eta_c$) two solutions appear. The one with larger value of
$c_-$ merges with the $\ell=0$ solution, as $c_-\to 1$ when
$\eta\to \lambda q/2$. This solution is un-physical as it
describes a network where $c_+$, and hence the average degree,
decreases with the networking effort $\eta$ (or with decreasing
volatility $\lambda$). Indeed, a detailed analysis of the linear
stability reveals the presence of an unstable mode \cite{footnote4}.

The lower branch instead has $c_-\to 0$ as $\eta\to\infty$ and it
describes a physical solution with average degree increasing with
$\eta/\lambda$. Numerical analysis of the stability matrix shows
that this branch is indeed dynamically stable.

The critical point $\eta_c(q)$ at which the $\ell=1$ solution
converges to $\lambda$ when $q\to 2$, which is the point
where the $\ell=0$ solution ceases to exist. So the transition is
continuous for $q=2$ and there is only one branch. For $q=10$ we
find $\eta_c = 2.27989...\lambda$ and for large $q$ we find
$\eta_c\propto \log q$.

In summary, the system has either one or two stable states,
depending on the values of $q$ and $\eta/\lambda$. For
$\eta<\eta_c$ only the solution $\ell = 0$ is stable, for
$\eta>\lambda q/2$ only the solution $\ell = 1$ is stable. Finally
in the interval $\eta_c<\eta<\lambda q/2$ there are two stable
solutions. The coexistence region $[\eta_c,\lambda q/2]$ shrinks
to a single point when $q=2$ and it gets larger as $q$ increases.

\begin{figure}
\includegraphics[width=0.9\columnwidth,clip]{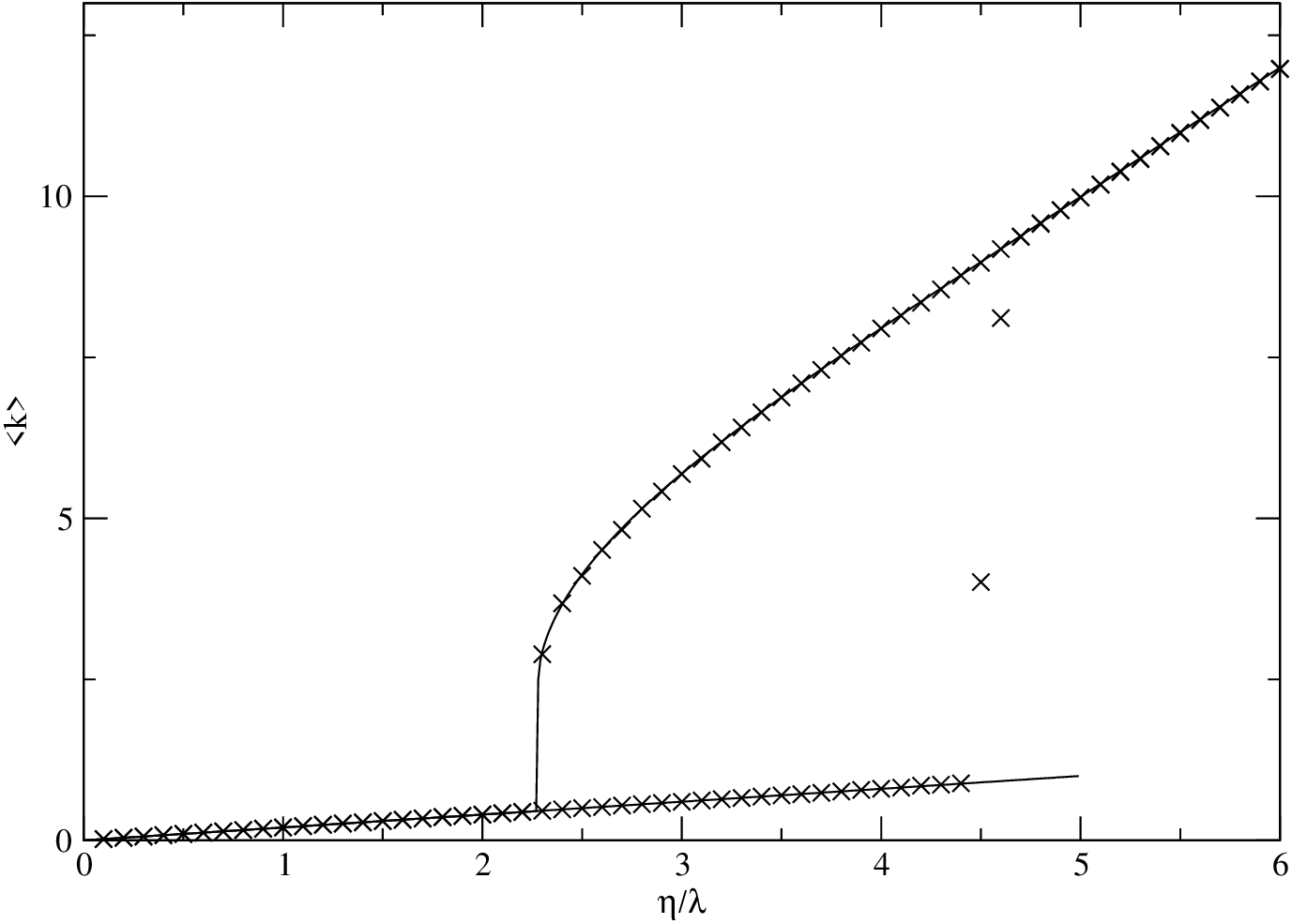}
\caption{
Plots of the mean connectivity $\left< k \right>$ as a function of $\eta / \lambda$.
Lines are theory, crosses are simulations, $n=10000$, run to $t=100$ for equilibration, then to $t=200$ for data taking.  For n=10000 the q=10 low state is unstable below the predicted value of 5 due to fluctuations being significant for finite n.  The crosses that do not lie on the theory curves are systems which 'jumped' during the data taking.
\label{crpfig}}
\end{figure}

Figure \ref{crpfig} shows plots of $\left< k \right>$ against
$\eta / \lambda$ for $q=10$ for both simulations and the theory
described here.

Concerning the degree distribution, it is worth noticing that the
$\ell = 0$ solution is characterized by a trivial Poisson
distribution for the whole random graph. Since $\left< k
\right>=c_- < 1$, there is no giant component and the system is
composed of many disconnected components of few nodes. The
solution with $\ell = 1$ is however non-trivial. In this case we
have a network whose $p(k)$ is the sum of two Poisson
distributions, one of which has $c_+ > 1$ and thus a giant
component, whilst the other (consisting of $q-1$ separate networks
plus the $k=0$ nodes) has $c_- < 1$ and thus has no giant
component.

\begin{figure}
\includegraphics[width=0.9\columnwidth,clip]{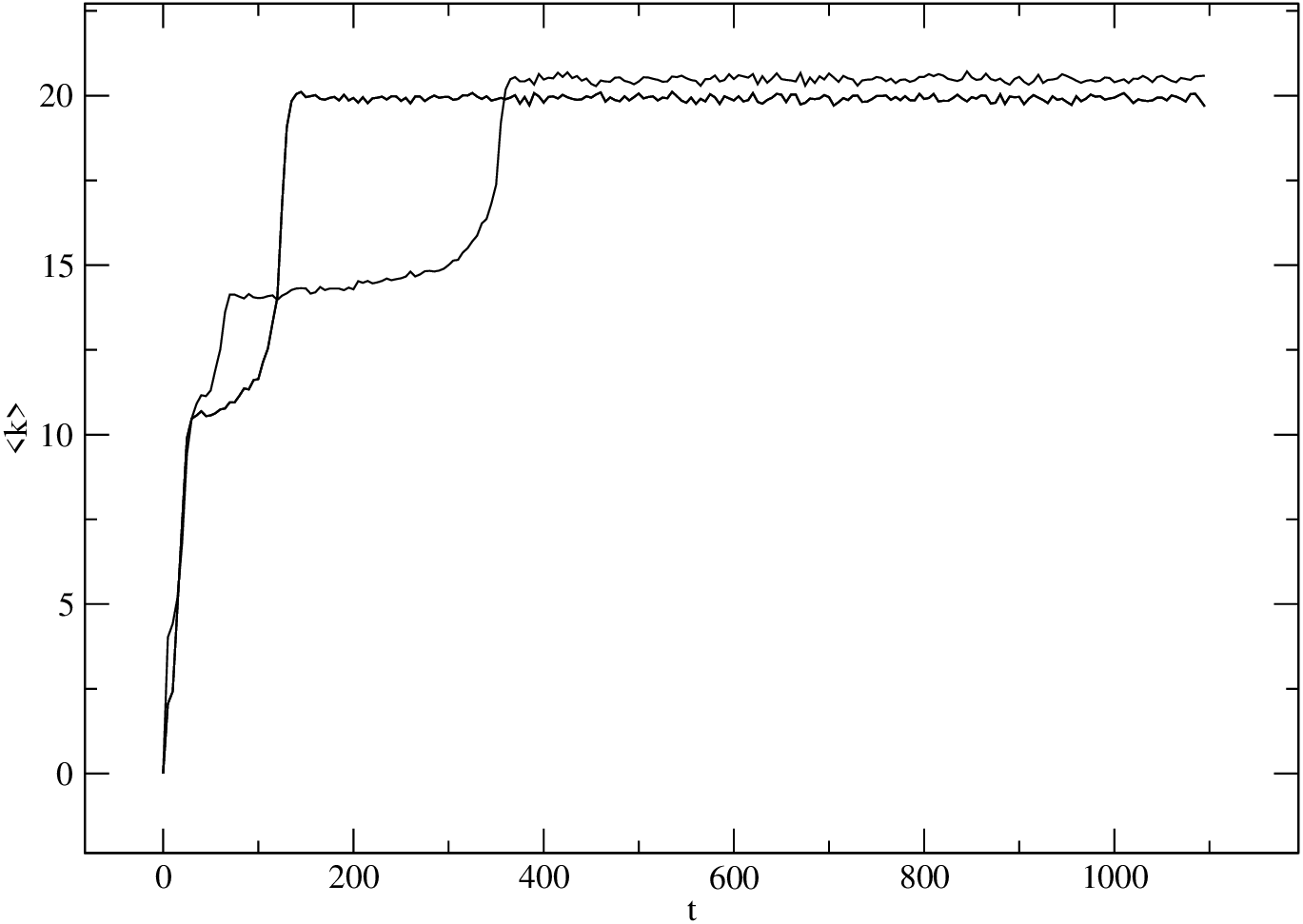}
\caption{ Plots of mean connectivity ${k}$ against time for networks starting in an initially unconnected state with $n=5000$ and $q=10$ and $\lambda=1$.  The curve on the top right is $\eta=20$ and the other is $\eta=10$.
\label{freezingfig}}
\end{figure}

Even if $\ell>1$ states are unstable, they may occur in the early
stages of the stochastic evolution of the network. Figure
\ref{freezingfig} shows time-series plots of simulations for
relatively large $\eta$ ($\eta=10,20$), starting in an initially
unconnected state. Although the $\eta=10$ solution eventually
reaches its expected value of $\left< k \right> \cong 20$, its
approach to that value is not smooth as might have been expected.
Rather we find that the network spends some time in an
intermediate $\ell>1$ metastable state. As we move to $\eta=20$,
the time spent in metastable states increases substantially,
failing to reach the stationary $\ell=1$ state (where $\left< k
\right> \cong 40$) despite the relatively long simulation time.
The reason for this behavior is that initially more than one giant
component forms, with different values of $\sigma$. This state
persists for a typical time $t_{\rm meta}$ which is inversely
proportional to the eigenvalue of the unstable mode. Hence $t_{\rm
meta}\sim 1/\mu\sim e^{c_+} \eta / (\lambda c_+^2)$ becomes very long
when $c_+$ is large. The reason why the dynamics is so slow
depends on the fact that in order for nodes to migrate from one
component $\sigma$ to another one, they have to loose all their
links. Such a process is limited by the density $p_{\sigma,1}$ of
nodes in component $\sigma$ with $k=1$, which is very small when
$c_+$ is large ($p_{\sigma,1}\sim e^{-c_\sigma}$).

In such a situation introducing a stochastic element in agents'
choice (i.e. switching on $T>0$) or allowing for the formation of
uncoordinated links (i.e. $\epsilon>0$) would make the system
converge very fast to the coordinated state. In other words, this
is a case where a finite ``temperature'' may allow the agents to
find the global optimum more quickly and it might be rational for
agents to resort to a stochastic choice rule. The ability to find
an optimal state more quickly is also of advantage if there are
external (exogenous) shocks which occasionally perturb the system.

\subsection{Discussion}

The $T=0$ case is of particular interest because it can be solved
exactly. For the other models described here, the
coordination/correlation of the nodes were too complex for us to
analyse exactly. In this section we have exactly solved a
non-trivial network model. This was possible because of the fact
that an agent only changes its spin if it has degree zero. The
network is found to be the sum of $q$ Poissonian random graphs
\cite{gecomment}.

\section{Conclusion}
\label{sec:concl}

In this paper we have proposed a general theoretical setup to
study the dynamics of a social network that is flexible enough to admit a
wide variety of particular specifications. We have studied three such
specifications, each illustrating a distinct way in which the network
dynamics may interact with the adjustment of node attributes. In all these
cases, network evolution displays the three features (sharp transitions,
resilience, and equilibrium co-existence) that empirical research has found
to be common to many social-network phenomena. Our analysis indicates that
these features arise as a consequence of the cumulative self-reinforcing
effects induced by the interplay of two complementary considerations. On the
one hand, there is the subprocess by which agent similarity is enhanced
across linked (or close-by) agents. On the other hand, there is the fact
that the formation of new links is much easier between similar agents. When
such a feedback process is triggered, it provides a powerful mechanism that
effectively offsets the link decay induced by volatility.

The similarity-based forces driving the dynamics of the model are
at work in many socio-economic environments. Thus, even though
fruitful economic interaction often requires that the agents
involved display some \textquotedblleft complementary
diversity\textquotedblright\ in certain dimensions (e.g. buyers
and sellers), a key prerequisite is also that agents can
coordinate in a number of other dimensions (e.g. technological
standards or trading conventions). Analogous considerations arise
as well in the evolution of many other social phenomena (e.g. the
burst of social pathologies discussed above) that, unlike what is
claimed e.g. by Crane \cite {Crane}, can hardly be understood as a
process of epidemic contagion on a \emph{given} network. It is by
now well understood \cite{Bailey, PV} that such epidemic processes
do \emph{not} match the phenomenology reported in empirical
research. Our model suggests that a satisfactory account of these
phenomena must aim at integrating both the dynamics \emph{on} the
network with that \emph{of} the network itself as part of a
genuinely co-evolutionary process.

One common feature of all the models discussed in this paper is
that stable states can have either a single giant component or
none. Many real situations are characterized by stable states with
a multitude of distinct components, barely connected. One example
is the polarization of opinion (e.g. in politics) where the
tendency of individuals to have opinions similar to those of the
peers they interact with may also lead to the segregation of the
population in different communities, of like mined individuals.
The results presented here suggest that there must be a specific
mechanism which is responsible for such a polarization. We hope
that future work in this direction may shed some light on this
issue.

\acknowledgments

Work supported in part by the European Community's Human Potential
Programme under contract HPRN-CT-2002-00319.


\begin{thebibliography}{99}
\bibitem{footnote1}  Specifically, Hagerdon \cite{Hagedoorn}
reports that 99\% of the R\&D partnerships worldwide are conducted
among firms in the so-called Triad: North America, Europe and Japan.
\bibitem{footnote2}  We recall that saying that a Poisson process occurs
at a rate $r$ is equivalent to saying that it occurs with
probability $r dt$, independently, in any infinitesimal time
interval $[t,t+dt)$.
\bibitem{footnote3}  In order to
  derive such an equation, fix a small time interval $dt$. If $\nu
  dt\gg 1$ the number of updates on each site will be large and hence,
  by the central limit theorem, the corresponding
  increments in the $h_i$'s are well approximated by a deterministic
  term equal to the expected value of the r.h.s. of Eq. (1), times
  $dt$, plus a random Gaussian contribution.
\bibitem{footnote4}  In this case there are $q-2$ eigenmodes describing mass transfer
across $\sigma>1$ components (i.e. $\epsilon_1=0,~\sum_\sigma\epsilon_\sigma=0$) and two eigenmodes with $\epsilon_1\neq 0$ and $\epsilon_\sigma=\epsilon_0$ for
$\sigma>1$. One of these is unstable.
\bibitem{gecomment}  {The fact that the sub-graphs are Poissonian (that they {\it are} fully specified by their degree distribution) is linked to the fact that we could find an exact solution.}
\end{thebibliography}
\end{document}